\documentclass[12pt,manuscript,psfig,epsfig,superscriptaddress]{revtex4}
\usepackage[latin9]{inputenc}
\usepackage{amsmath}
\usepackage{amssymb}
\usepackage{graphicx}

\makeatletter
\@ifundefined{textcolor}{}
{%
 \definecolor{BLACK}{gray}{0}
 \definecolor{WHITE}{gray}{1}
 \definecolor{RED}{rgb}{1,0,0}
 \definecolor{GREEN}{rgb}{0,1,0}
 \definecolor{BLUE}{rgb}{0,0,1}
 \definecolor{CYAN}{cmyk}{1,0,0,0}
 \definecolor{MAGENTA}{cmyk}{0,1,0,0}
 \definecolor{YELLOW}{cmyk}{0,0,1,0}
}

\usepackage{dcolumn}
\usepackage{bm}
\usepackage{amsfonts}

\makeatother

\begin{document}

\title{Color connections of the four-quark $Q\bar{Q}Q'\bar{Q}'$ system and
doubly heavy baryon production in $e^{+}e^{-}$ annihilation}

\author{Yi Jin}

\affiliation{School of Physics and Technology, University of Jinan, Jinan 250022,
P. R. China}

\author{Shi-Yuan Li}

\affiliation{School of Physics, Shandong University, Jinan 250100, P. R. China}

\author{Zong-Guo Si}

\affiliation{School of Physics, Shandong University, Jinan 250100, P. R. China}

\author{Zhong-Juan Yang}

\affiliation{School of Physics and Chemistry, Henan Polytechnic University, Jiaozuo 454003, P. R. China}

\author{ Tao Yao}

\affiliation{School of Physics, Shandong University, Jinan 250100, P. R. China}
\begin{abstract}
The hadronization effects induced by various color connections of
the four-quark system in $e^{+}e^{-}$ annihilation are briefly reviewed.
A special color connection of the four-heavy-quark system with no color separation
favors the production of doubly heavy baryons. For the related three-jet
case, the corresponding hadronization has not yet been considered. We
argue that it can be effectively described as two-string fragmentation
in addition to the leading heavy diquark fragmentation. The production rate
and the properties of the final-state hadron systems are discussed. Emphasis
is placed on the string effect as a fingerprint of this hadronization procedure.


\end{abstract}
\maketitle

\section{Introduction}

The hadronization of the parton system created in hard collisions
is very crucial to the understanding of the confinement mechanism, and it is very
important to directly compare to data for a theoretical prediction
with strongly interacting final states. However, up to now, it has not been possible to describe the hadronization process perturbatively, and it has been necessary to deal with
various hadronization models. The color connection of the final parton
system is necessary to set the `surface' between the Perturbative
Quantum Chromodynamics (PQCD) phase and the hadronization phase, which is beyond the approach of PQCD.

It has been argued that some clues can be obtained by analyzing the
decomposition of the color space of the final partons produced in
the PQCD phase \cite{Gustafson,xie1,xie2}. For the color space of the
parton system, there are many possible decompositions that correspond to
various color connections (see, \textit{e.g.}, \cite{WQEFH,Wagus}).
Hence, it is instructive to study the special properties
of the final hadrons, which could provide information regarding evidence
of specific color connections. Such investigations
have been conducted for various parton systems \cite{Friberg:1996xc,xie3,LEP2}.
One interesting and important example is the four-quark system ($q_{1}$
$\bar{q}_{2}$ $q_{3}$ $\bar{q}_{4}$) produced via hard collisions.
This is the simplest system in which many phenomena/mechanisms related
to QCD properties can be studied, \textit{e.g.}, the (re)combination of
quarks in the production of special hadrons, the influence of soft interaction on the reconstruction of intermediate particles (such as $W^{\pm}$), \textit{etc.},
most of which are more or less related to the color connections among
these four quarks.

In our previous work \cite{hlsy}, we have noted that the hadronization
effects of one kind of the color connections of such four-quark systems
($q_{1}$ $\bar{q}_{2}$ $q_{3}$ $\bar{q}_{4}$) had not yet been
investigated. In this kind of color connection, these four quarks
cannot be grouped into two singlet clusters. They can
only be treated as a whole for hadronization. The special case (diquark pair
fragmentation) in $e^{+}e^{-}$ annihilation has been investigated
in ref. \cite{hlsy}. The trigger there is the two-jet event shape,
which corresponds to the kinematical configuration in which both $(q_{1}q_{3})$
and $(\bar{q}_{2}\bar{q}_{4})$ have small invariant masses and can
be treated as a diquark and an antidiquark, respectively. This leads to the production of two baryons
as the leading particles of these two jets.

However, the above kinematical restriction is very strong for that
special case. On the other hand, to trigger such kind of color connection,
it is feasible to employ the hadrons containing two heavy quarks from
the doubly heavy diquark fragmentation. Here, we investigate the four-heavy-quark system in which only one $QQ'$ (or $\bar{Q}\bar{Q}'$) pair
has a small invariant mass and can be considered as a diquark, while the invariant masses of the other two are left unrestricted. To clearly describe such a configuration, we
consider a three-jet event shape. The doubly heavy diquark can
fragment not only into various doubly heavy baryons (\textit{e.g.},
$\Xi_{cc}$ \cite{Ma:2003zk, Jiang:2013ej}) but also into doubly heavy tetraquarks (\textit{e.g.}, $T_{cc}$ \cite{liuyanrui}), depending on whether the doubly heavy diquark combines with a quark or an antidiquark. In this paper, the case of tetraquark production will also be briefly addressed when necessary to exhaust all the possible hadronic final states.

In future $Z^{0}$ factory or Giga-Z programs in high-luminosity next-generation $e^+e^-$ high-energy colliders (see, \textit{e.g.}, \cite{zfactory}), it will be possible to thoroughly experimentally study such a special color configuration. We set the
center-of-mass energy to the $Z^{0}$ mass as an example for the feasibility
of the numerical calculations, and at this energy, the heavy quarks Q/Q' refer to the charm and/or bottom quarks.

This paper is organized as follows. In section 2, we review the color
connections and the corresponding hadronizations of the four-quark
system. In section 3, we analyze the event-shape and phase-space
configurations at the parton level for the case in which we are interested. In section 4, we describe the hadronization
of the special color connection and present the numerical results. Finally,
we provide a short summary.

\section{Color connections of the four-quark system}

For the $q_{1}\bar{q}_{2}q_{3}\bar{q}_{4}$ system, two of the various decompositions of its color space can be written as follows: 
\begin{equation}
(3_{1}\otimes3_{2}^{*})\otimes(3_{3}\otimes3_{4}^{*})=(1_{12}\oplus8_{12})\otimes(1_{34}\oplus8_{34})=(1_{12}\otimes1_{34})\oplus(8_{12}\otimes8_{34})\oplus\cdots,\label{1stdp}
\end{equation}
\begin{equation}
(3_{1}\otimes3_{4}^{*})\otimes(3_{3}\otimes3_{2}^{*})=(1_{14}\oplus8_{14})\otimes(1_{32}\oplus8_{32})=(1_{14}\otimes1_{32})\oplus(8_{14}\otimes8_{32})\oplus\cdots,\label{2nddp}
\end{equation}
where $3$, $3^{*}$, $1$ and $8$ denote the
triplet, anti-triplet, singlet and octet representations, respectively, of the $SU_{c}(3)$ Group, and the subscripts correspond to the relevant (anti)quarks. In these
two color decompositions, the quark and antiquark form a cluster/string,
which fragments into hadrons independently. This picture is adopted
in the popular hadronization models (for details, see \cite{string,pythia,cluster,herwig}).
The difference between these two decompositions leads to the color
reconnection phenomena, which are relevant to processes such as $e^{+}e^{-}\to W^{+}W^{-}/Z^{0}Z^{0}\to q_{1}\bar{q}_{2}q_{3}\bar{q}_{4}\to4~jets$
at LEPII \cite{LEP2,khojos,gustafsonw,Gustafson:1994cd,xie1,lsya},
\textit{etc}. One might also notice that in the $e^{+}e^{-}\to q_{1}\bar{q}_{1}+g^{*}\to q_{1}\bar{q}_{1}q_{2}\bar{q}_{2}$
process, ($q_{1}\bar{q}_{1}$) and ($q_{2}\bar{q}_{2}$) cannot exist
in color singlets, while ($q_{1}\bar{q}_{2}$) and ($q_{2}\bar{q}_{1}$)
are usually treated as color singlets and hadronize independently.
This corresponds to the term $1_{14}\otimes1_{32}$ in Eq. (\ref{2nddp}) \cite{xie4}.

One of the important properties of the above decompositions is
that the entire system is decomposed into color-singlet clusters.
As noted in \cite{hlsy}, the color space of the $q_{1}\bar{q}_{2}q_{3}\bar{q}_{4}$
system can also be decomposed in other ways to construct the color-singlet state as whole. Here, we are interested in the case in which 
\begin{equation}
(3_{1}\otimes3_{3})\otimes(3_{2}^{*}\otimes3_{4}^{*})=(3_{13}^{*}\oplus6_{13})\otimes(3_{24}\oplus6_{24}^{*})=(3_{13}^{*}\otimes3_{24})\oplus(6_{13}\otimes6_{24}^{*})\oplus\cdots,\label{diqcl}
\end{equation}
where $6_{13}$ ($6_{24}^{*}$) denotes the sextet
(anti-sextet) representation of the $SU_{c}(3)$ Group. When two (anti)quarks in
color state $3^{*}(3)$ attract each other and form a(n) `(anti)diquark'
and their invariant mass is sufficiently small, such a cluster has a certain probability
of hadronizing like a(n) (anti)diquark, as discussed in our previous work,
by triggering the leading baryons and two-jet-like event shape. For
the case in which only one of the pair has a small invariant mass, the color
configuration can be better written as 
\begin{eqnarray}
(3_{1}\otimes3_{3})\otimes3_{2}^{*}\otimes3_{4}^{*}=3_{13}^{*}\otimes3_{2}^{*}\otimes3_{4}^{*}\oplus\cdots, & or\nonumber \\
3_{1}\otimes3_{3}\otimes(3_{2}^{*}\otimes3_{4}^{*})=3_{1}\otimes3_{3}\otimes3_{24}\oplus\cdots.\label{q2qq}
\end{eqnarray}
The color configuration as a whole is like a `big baryon.' Without a
doubt, the (anti)quark pair with a small invariant mass can hadronize
into a(n) (anti)baryon (tetraquark) as a(n) (anti)diquark. As mentioned above,
when these four quarks are all heavy, it is easy to identify the doubly
heavy baryon/tetraquark, which comes directly from the doubly heavy
diquark. As an example, for the $cc\bar{c}\bar{c}$ system, we can
trigger the $\Xi{}_{cc}$/$T_{cc}$. In this case, the diquark $cc$ must combine
with a quark $q$ (antidiquark) to hadronize into the baryon (tetraquark), which leaves the $\bar{q}$ (diquark) to compensate the quantum number of the system (hereafter, $q$ represents a light quark). For the remaining system, to describe the hadronization,
a concrete model must be assigned, which will be investigated in detail in this paper.

Compared with our previous work concerning two-jet-like events \cite{hlsy} in which the color connection is the same as that of Eq. (\ref{diqcl}), the phase space of this three-jet event is much larger and leads to a larger cross section.  
In this kind of events, the doubly heavy (anti)diquark belongs to an isolated jet separated from the other two jets induced by two heavy antiquarks (quarks).
As a result, it is easy to trigger four-heavy-quark final states for the sake of studying their color connections.
In addition, corresponding to the color connection in Eqs. (\ref{1stdp}) and (\ref{2nddp}), two heavy quark-antiquark ($Q\bar{Q}$) pairs form
two clusters. If the invariant mass of the $Q\bar{Q}$ cluster is within the proper range, it will fragment into heavy quarkonia or 
tetraquark states, \textit{etc}. Recently, a large number of new bound states, \textit{e.g.}, X, Y and Z particles, have been observed in experiments. 
It has long been suggested that the production mechanism can help to identify the 
structure of the hadrons \cite{Maiani:2006ia, Han:2009jw}.
One important example is X(3872). Its production mechanism has been carefully studied in \cite{Bignamini:2009sk, Esposito:2013ada} to explore its structure, and as a result, X(3872) has been interpreted as a hadron molecule. However, X(3872) may also be interpreted as a tetraquark, $c q  \bar{c}\bar{q}$. For this case, within our framework, if the invariant mass of the $c\bar{c}$ cluster is within the proper range, X(3872) can be fragmented from the $c\bar{c}$ cluster corresponding to the color connections of Eqs. (\ref{1stdp}) and (\ref{2nddp}), \textit{i.e.}, $c\bar{c}\to X(3872)+\cdot\cdot\cdot$. Without detailed investigation, we can roughly estimate the cross section for the prompt
production of X(3872) by demanding that the invariant mass of the $c \bar{c}$ cluster be slightly larger than $M_{X(3872)}$, \textit{i.e.}, $M_{c\bar{c}}=M_{X(3872)}+\delta m $ at the $Z^0$ pole energy in $e^+e^-$ annihilation. The result is approximately $1.5 \times 10^{-2}$ pb for $\delta m=3.0$ GeV
($4.5 \times 10^{-3}$ pb for $\delta m=1.5$ GeV). 

The aim of this work is to investigate the hadronization effects corresponding to Eq. (\ref{q2qq}) that are related to the production of $\Xi_{cc}$ and $T_{cc}$. Therefore, in the following, we will focus on investigating the
effects of the events that contain $\Xi_{cc}$ or $T_{cc}$ that may be observed in experiments at $e^+e^-$ colliders at the $Z^0$ pole energy. Before employing the model to
calculate the final hadron events, the three-jet event
shape must be investigated, and this is discussed in the next section. It is independent of the hadronization and can act as a further trigger to eliminate the ambiguities of the special color connection.

\section{Event-shape analysis at the parton level}

\begin{figure}[htb]
\centering 
\scalebox{0.4}{\includegraphics{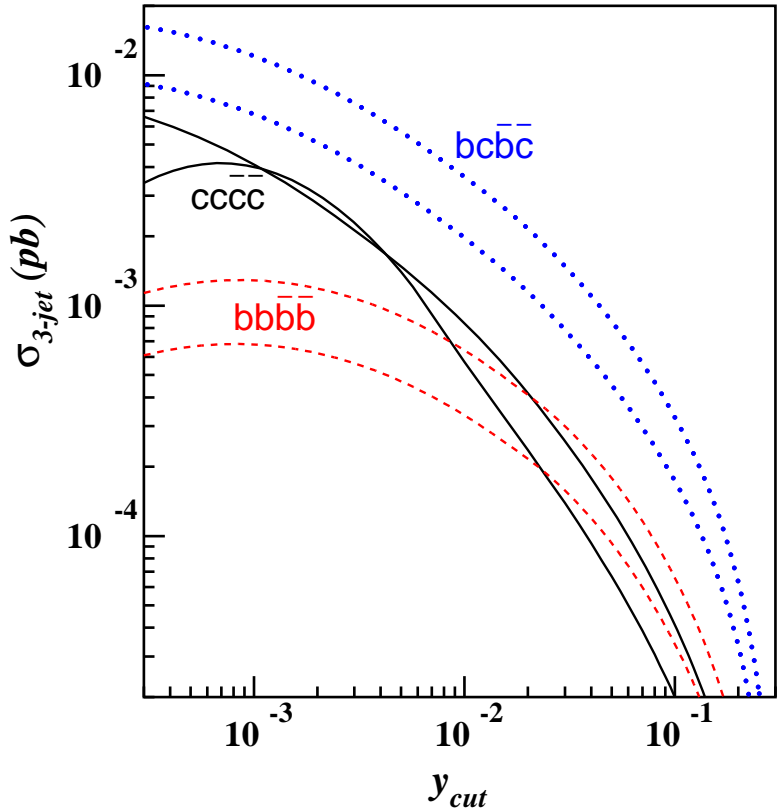}} 
\caption{The three-jet cross section $\sigma_{3-jet}$ with respect to $y_{cut}$ for
$e^+ e^- \to Q Q' \bar{Q} \bar{Q}'$. The higher solid/dashed/dotted line represents $\delta m$ = 1.5 GeV, and the corresponding lower one represents $\delta m$ = 1.0 GeV.}
\label{4csig} 
\end{figure}

The differential cross section for $e^{+}e^{-}\to Q\bar{Q}Q'\bar{Q}'$
can be written as 
\begin{equation}
d\sigma=\frac{1}{2s}d\mathcal{L}ips_{4}\overline{|\mathcal{M}|^{2}},\label{sigma1}
\end{equation}
where $s$ is the square of the total energy, ${\cal L}ips_{4}$ represents
the 4-particle phase space, and $\overline{|\mathcal{M}|^{2}}$ is
the spin average and summation of the squared modulus amplitude. For this process,
at the energy of the $Z^{0}$ pole, we investigate the three-jet cross
section. Here, we require that the invariant mass of $QQ'$ or $\bar{Q}\bar{Q}'$
lie within the range between $M_{Q}+M_{Q'}$ and $M_{Q}+M_{Q'}+\delta m$.
As is well known, a well-defined jet is infrared safe, and the parton-level result can be applied at the hadron level. Here, we take the Durham
algorithm \cite{DURHAM} 
\begin{equation}
y_{ij}=\frac{2min(E_{i}^{2},E_{j}^{2})(1-cos\theta_{ij})}{E_{cm}^{2}}\label{cutoff-1}
\end{equation}
to define the jet. The parameter $y_{cut}$ is thus introduced, and two particles are considered as being in one jet when $y_{ij}<y_{cut}$. We apply this jet algorithm to the partonic cross section, 
\begin{equation}
d\tilde{\sigma}=d\sigma[\Theta(M_{Q}+M_{Q'}+\delta m-M_{QQ'})+\Theta(M_{\bar{Q}}+M_{\bar{Q}'}+\delta m-M_{\bar{Q}\bar{Q}'})],
\end{equation}
to obtain the three-jet cross section $\sigma_{3-jet}$. In the above
equation, $M_{QQ'}(M_{\bar{Q}\bar{Q}'})$ represents the (anti)diquark
invariant mass. In our calculation, we take the fine structure constant to be
$\alpha=1/128$ and the strong coupling constant to be $\alpha_{s}$ = 0.12.
As an example, we set the quark masses to be $m_{c}$ = 1.5 GeV/$c^{2}$ and
$m_{b}$ = 4.5 GeV/$c^{2}$. The corresponding results
for the three-jet cross section $\sigma_{3-jet}$ are shown in Fig. \ref{4csig} for $\delta m$=1.5 GeV and 1.0 GeV.

It is easy to find that the three-jet cross section can reach the order of 
$10^{-2}$ pb for $y_{cut} = 10^{-3}$. If the integrated luminosity of a $Z^{0}$ factory is approximately $10^{5}$ pb$^{-1}$, then the number of heavy-diquark production events can be on the order of one thousand. This makes it possible
to investigate the hadronization effects of the special color connection of Eq. (\ref{q2qq}).

In the above calculations, the weight of $(3^*_{13}\otimes 3_{24})$ relative to $(6_{13}\otimes 6^*_{24})$ is not included.  
If only perturbative QCD is considered, this relative weight can be easily calculated by employing the color effective Hamiltonian \cite{cs} for this process:
\begin{equation}\label{hc}
H_c= \Big[ \big (\Psi_{1i}^\dag \Psi^{2j\dag}-\delta_i^j\Psi_{1k}^\dag \Psi^{2k\dag}/3 \big ) \cdot 
\big (\Psi_{3j}^\dag \Psi^{4i\dag}-\delta_j^i\Psi_{3k'}^\dag \Psi^{4k'\dag}/3 \big ) \Big ] \cdot D,
\end{equation}
where $D$ represents the phase-space factor. Then, 
\begin{equation}
prob(|3_{13}^{*}\otimes3_{24} \rangle ) = 
\Big | \big \langle 0 | 1/6 \cdot \epsilon_{mnp}a_{1n}a_{3p} \cdot \epsilon_{m'n'p'}b_{2n'}b_{4p'} 
\cdot H_c | 0 \big \rangle \Big |^2 =\frac{4}{9}.
\label{prob}
\end{equation}
Hence, the relative weight of $(6_{13}\otimes 6^*_{24})$ is 5/9. 
This result, however, is not strongly predictive because non-perturbative QCD determines the probability of each decomposition. 
In this case, the small $QQ'$ mass may enhance the probability of the decomposition represented by Eq. (\ref{diqcl}). 
This crucial probability of each decomposition must be determined from data, as discussed here.

\section{Hadronization and results}

For a color-singlet system with a large invariant mass, its hadronization
is a `branching' process via the creation of quarks from the vacuum by
the strong interactions within the system. The created quarks and the
primary quarks are combined into color-singlet hadrons. Here, as an example, we start from a heavy diquark.
The heavy diquark needs a quark/antidiquark to form an open doubly
heavy baryon/tetraquark. To balance the quantum numbers of color
and flavor, an antiquark/diquark must be simultaneously created
from the vacuum. To branch them further, more quark pairs and diquark
pairs must be created from the vacuum via the interactions among the quark
system. Such a cascade process will proceed until the end of time,
when most of the `inner energy' of the entire system is transformed into
the kinematical energies and masses of the produced hadrons. Each
of two newly created quarks (antidiquarks) combines with each of the
primary heavy antiquarks to respectively form two open heavy-flavor
hadrons.

\begin{figure}[htb]
\centering 
\scalebox{0.4}{\includegraphics{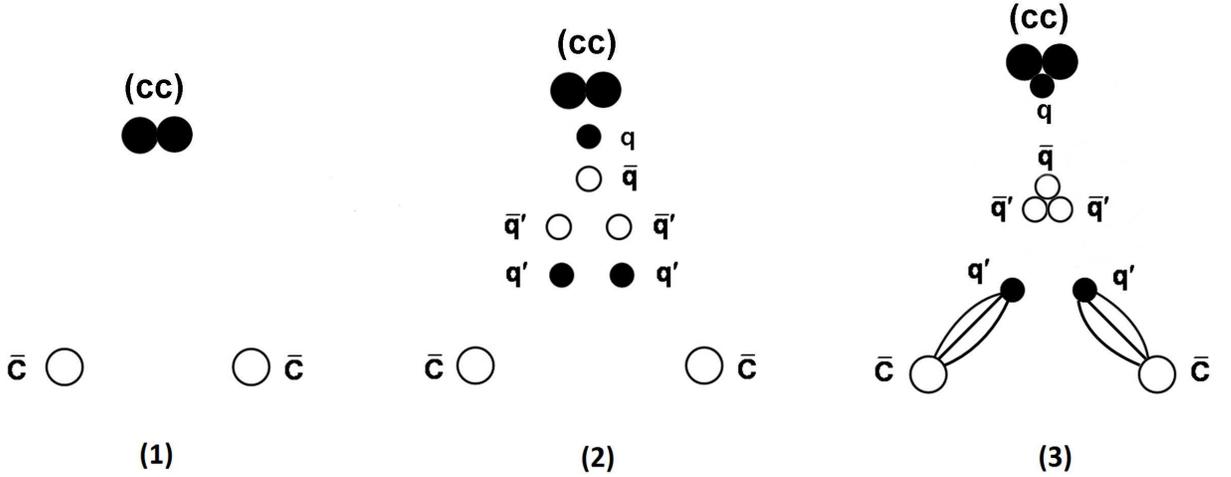}} 
\caption{The string formation for the hadronization of the $(cc')\bar{c}\bar{c}'$
system with the aid of quark creation from the vacuum. Solid circles represent quarks, while hollow circles represent antiquarks. The primary $\bar c$ and $\bar c$ connect to quarks respectively via two strings in (3).}
\label{4cfrag} 
\end{figure}

For the configurations considered here, the above process is straightforward,
except that for each step, we must assign special quantum numbers
for each specific kind of hadron according to its production rate.
Because of the success of the Lund string model, especially its realization
by PYTHIA/JETSET, the above hadronization procedure can be easily
realized. For the case of tetraquark production, if the complementary
diquark pair is broken by the interactions within the remaining system
and then each becomes connected to the other two primary heavy antiquarks to form
two strings, the resultant hadronization can be described by the conventional
string-fragmentation picture. For the case of doubly heavy baryon production, the complementary antiquark can produce a baryon by combining with an antidiquark, and then the balancing diquark can help to form two strings in the same manner described above.
This procedure is illustrated in Fig. \ref{4cfrag}, where the $(cc)\bar c\bar c$ system and the $\Xi_{cc}$ production are used as an example.
 
The fragmentation of the heavy diquark is described by the Peterson
formula \cite{pet83} 
\begin{equation}
f(z)\propto\frac{1}{z(1-1/z-\epsilon_{Q}/(1-z))^{2}},\label{Peterson}
\end{equation}
where $\epsilon_{Q}$ is a free parameter, which is expected to scale between
flavors as $\epsilon_{Q}\propto1/m_{Q}^{2}$. In the following, we will show two sets of results corresponding to $\epsilon_{Q}=1/25$ and $1/16$ for the $cc$ diquark. 

We will also try the modified Lund symmetric fragmentation function 
\begin{equation}
f(z)\propto\frac{1}{z^{1+r_Qbm_Q^2}}z^{a_\alpha}(\frac{1-z}{z})^{a_\beta}
\mbox{exp}(-\frac{bm_\bot^2}{z})
\label{modified}
\end{equation}
for heavy endpoint quarks, according to the Bowler (Artru\textendash Mennessier, Morris) space\textendash time picture of string evolution \cite{Bow81, Art74, Mor89}, to describe the fragmentation of the heavy diquark. This function is not sensitive to the free parameters $a_\alpha$, $a_\beta$ and $b$, but it is sensitive to $r_Q$. Here, we take $r_Q$=1.0, $a_\alpha=a_\beta=0.8$ and $b=0.58$, which are the default values in PYTHIA \cite{pythia}. 

To describe the fragmentation of the complementary (anti)quark, we adopt the fragmentation function employed by the LUND Group \cite{and83a} 
\begin{equation}
f(z)\propto z^{-1}(1-z)^{a}exp(-bm_{\perp}^{2}/z),\label{Anderson}
\end{equation}
where $a$ and $b$ are free parameters. In our program, we take $a=0.3$ GeV$^{-2}$
and $b=0.58$ GeV$^{-2}$, as used in PYTHIA \cite{pythia}. Another topic that we do not discuss here in detail is the excited states of the doubly heavy hadrons. If we consider the fact that heavy quark masses fatally break the SU(4) and/or SU(5)
flavor symmetries and we assume that all the excited states dominantly
decay into the ground state, the details of the differences can
be neglected. In the following, we provide the numerical results for $\Xi_{cc}$ and $T_{cc}$. The fragmentation of the strings can be referred to
the classical book about the Lund model by B. Anderson \cite{string} and the PYTHIA manual \cite{pythia}. The corresponding FORTRAN code contains three parts: The first part is the perturbative calculation codes employed in section 3 
to give the weight of each phase-space configuration of the primary quark system, which has been employed in our previous works \cite{Li:1999iz,Li:1999ar,Han:2009jw}. The second part describes the picture shown in Fig. \ref{4cfrag} and prepares the sub-strings, which are hadronized with the aid of PYTHIA in the third part. The FORTRAN code for the first two parts is available upon request, and the PYTHIA package can be found at its homepage.

\begin{figure}[htb]
\centering 
\scalebox{0.4}{\includegraphics{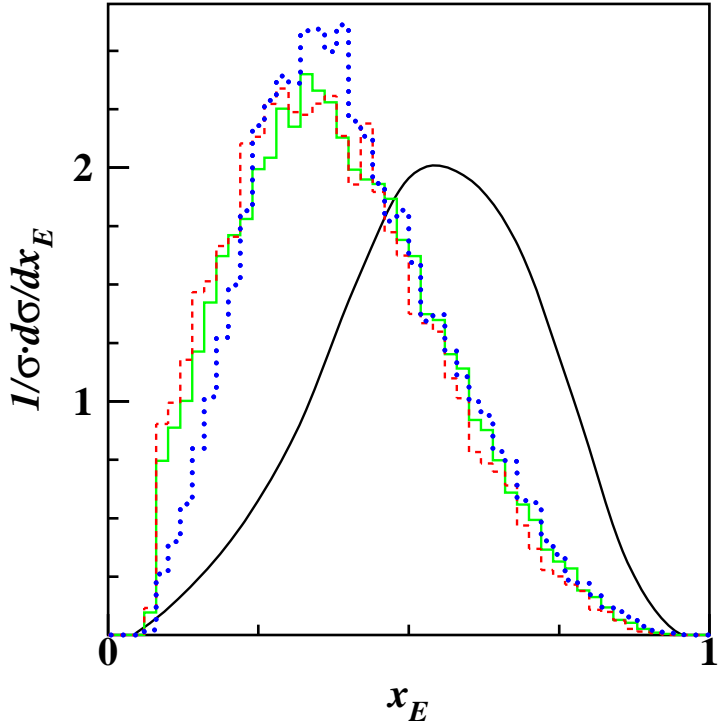}} 
\caption{The distribution of the energy fraction for $\Xi_{cc}$ compared to the diquark (cc) as a function of the scaled dimensionless variable $x_{E}=2E/\sqrt{s}$. Hereafter, three-jet events are defined with $y_{cut}=10^{-3}$ if no explicit explanation is given. The smooth, solid black line represents the quark pair of $cc$, the solid green line (dashed red line) represents the baryon $ccq$ that we trigger for $\epsilon_{Q}$=1/25 (1/16) in the Peterson formula Eq. (\ref{Peterson}), and the dotted blue line represents the modified Lund symmetric fragmentation function with $r_Q$=1.0 in Eq. (\ref{modified}).}
\label{xe} 
\end{figure}
\begin{figure}[htb]
\centering 
\scalebox{0.4}{\includegraphics{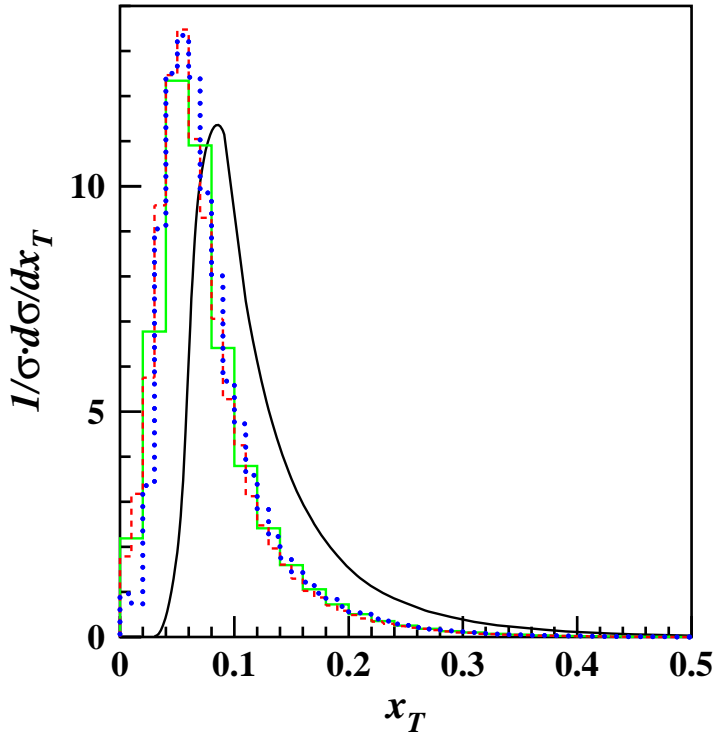}} 
\caption{The transverse-momentum distribution for $\Xi_{cc}$ compared to the diquark (cc) as a function of the scaled dimensionless variable $x_{T}=2p_T/\sqrt{s}$. The smooth, solid black line represents the quark pair of $cc$, the solid green line (dashed red line) represents the baryon $ccq$ that we trigger for $\epsilon_{Q}$=1/25 (1/16) in the Peterson formula Eq. (\ref{Peterson}), and the dotted blue line represents the modified Lund symmetric fragmentation function with $r_Q$=1.0 in Eq. (\ref{modified}).}
\label{xpt} 
\end{figure}
\begin{figure}[htb]
\centering 
\scalebox{0.4}{\includegraphics{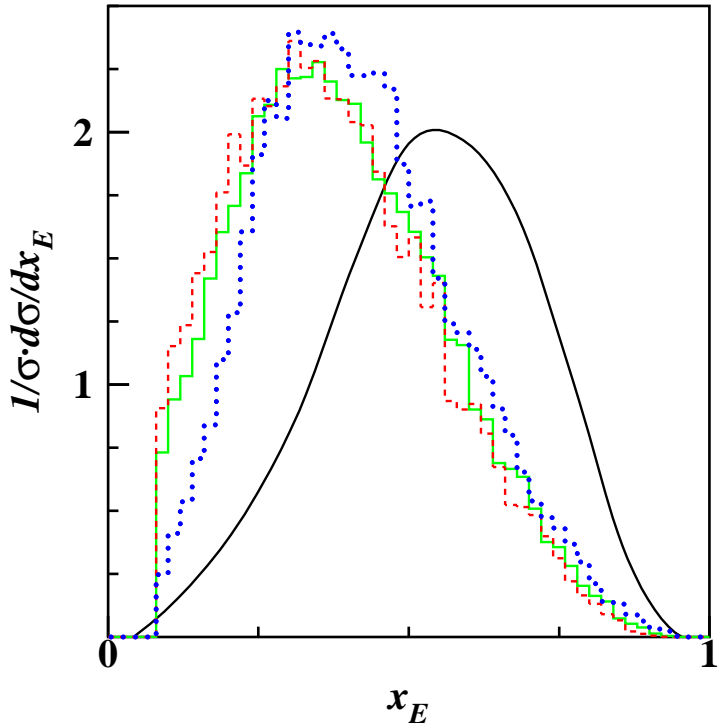}} 
\caption{The distribution of the energy fraction for $T_{cc}$ compared to the diquark (cc) as a function of $x_{E}=2E/\sqrt{s}$. The smooth, solid black smooth line represents the quark pair of $cc$, the solid green line (dashed red line) represents $T_{cc}$ for $\epsilon_{Q}$=1/25 (1/16) in the Peterson formula Eq. (\ref{Peterson}), and the dotted blue line represents the modified Lund symmetric fragmentation function with $r_Q$=1.0 in Eq. (\ref{modified}).}
\label{xetcc} 
\end{figure}
\begin{figure}[htb]
\centering 
\scalebox{0.4}{\includegraphics{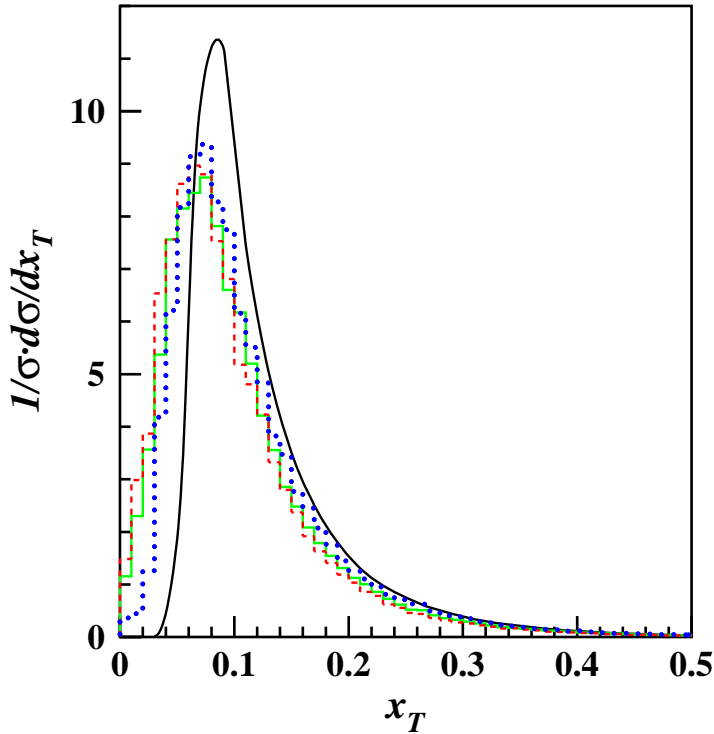}} 
\caption{The transverse-momentum distribution for $T_{cc}$ compared to the diquark (cc) as a function of $x_{T}=2p_T/\sqrt{s}$. The smooth, solid black line represents the quark pair of $cc$, the solid green line (dashed red line) represents $T_{cc}$ for $\epsilon_{Q}$=1/25 (1/16) in the Peterson formula Eq. (\ref{Peterson}), and the dotted blue line represents the modified Lund symmetric fragmentation function with $r_Q$=1.0 in Eq. (\ref{modified}).}
\label{xpttcc} 
\end{figure}

We show the energy and transverse-momentum (with respect to the thrust axis) distributions of $\Xi_{cc}$ and $T_{cc}$ in Figs. \ref{xe}, \ref{xpt}, \ref{xetcc} and \ref{xpttcc} respectively to demonstrate the hadronization effects. The absolute values of the distributions more or less depend on the parameters in Eqs. (\ref{Peterson},\ref{modified},\ref{Anderson}). The fragmentation functions and/or the parameters can be tuned by comparison to data once data are obtained. As seen from the energy-fraction distributions, one of the most significant hadronization effects can be observed in the fragmenting process. $\Xi_{cc}$ and $T_{cc}$ only take part of the energy of the $(cc)$, according to Eq. (\ref{Peterson}) or (\ref{modified}). From the transverse-momentum distributions, one can conclude that the transverse momentum (with respect to the thrust axis) of $\Xi_{cc}/T_{cc}$ that is created in the hadronization process is very small. We notice that these two kinds of hadrons have similar distributions. The difference for $x_T$ is more clear because the thrust is dependent on the entire hadronization, not only the diquark fragmentation.

\begin{figure}[htb]
\centering 
\scalebox{0.4}{\includegraphics{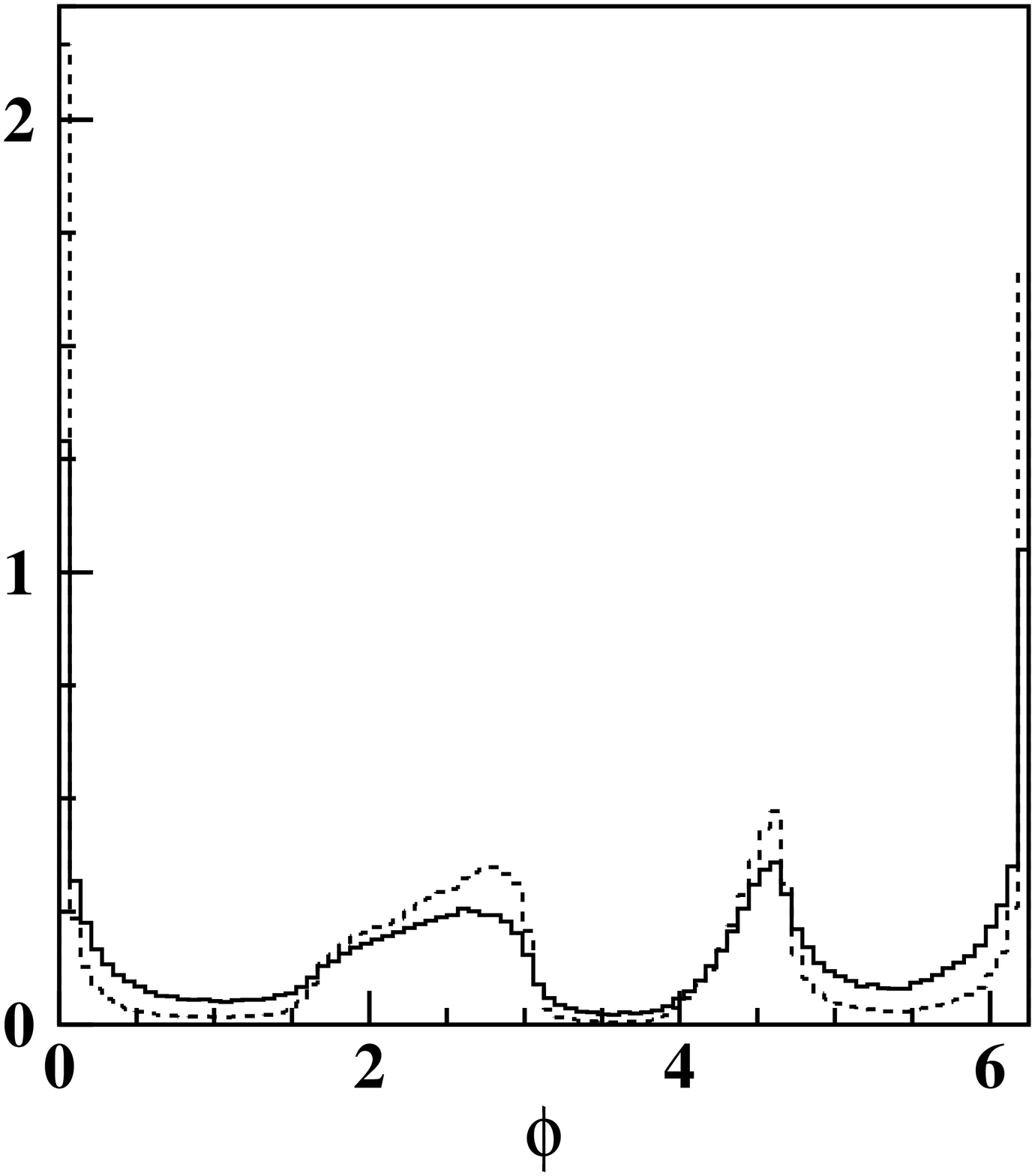}} 
\caption{The string effect in the three-jet case for $e^{+}e^{-}\to\Xi_{cc}+X$
events. The solid line represents $\frac{1}{N}\frac{dN}{d\phi}$, and the dashed line represents $\frac{1}{E}\frac{dE}{d\phi}$. Here, the symmetric three-jet events are defined to require each relative angle to be larger than $\frac{\pi}{2}$.}
\label{hnphi} 
\end{figure}

We would like to emphasize that the above description of the formation of
these two strings is only one of the effective descriptions of the branching procedure of the quark system. In principle, this and other effective models can yield the
same event-shape variables. Thus, these two strings can form after more
quark pairs and diquark pairs have been produced and combined, not necessarily
only after one step. However, the three-jet event shape is not sensitive
to at which step the strings form and fragment, nor is the leading doubly heavy hadron in one jet. This observation also applies 
to the following interesting phenomenon associated with such
a two-string configuration.
As seen from Fig. \ref{4cfrag}, because there are only two strings
rather than three, no string extends in the phase space between $\bar{Q}$
and $\bar{Q}'$. This leads to a significant string
effect, which is the fingerprint of the effective description.

As is well known, in the $e^{+}e^{-}$ center-of-mass frame, all three
jet momenta must be in the same plane (${\cal P}$) because of momentum
conservation \cite{Ellis:1976uc}, so it is straightforward
to observe the string effect \cite{stringeff}. Here, we provide a numerical
example. We choose the more symmetric three-jet events by requiring
that the angle between any two jets is larger than $\pi/2$. Then, the three-momentum
of each final-state particle $\vec{k}_i$ is projected onto one of the three regions between the jets to obtain $\vec{k}_i'$ in the plane ${\cal P}$. 
The three-momentum of the jet that contains the doubly heavy baryon is chosen to be the x axis. The angle between $\vec{k}_i'$ and the x axis is the azimuthal angle $\phi$ of 
the corresponding particle. 
We can then calculate the final particle-number (energy) distribution $1/N\,dN/d\phi$ ($1/E\,dE/d\phi$). 
The corresponding results are displayed in Fig. \ref{hnphi}.
Obviously, $\phi$=0 and $\phi=2\pi$ correspond to the momentum of
the jet that contains $\Xi_{cc}$. The region of $\phi$ around $\pi$
corresponds to the region with no string expansion, and few particles
emerge in this region. The peaks in Fig. \ref{hnphi} correspond to
the directions of the three jets. From this figure, we find that the particle-number distribution and the energy distribution are quite similar, so the string effect can be observed either via the particle number measured with a tracker spectrometer or via the energy measured with a calorimeter.

\begin{figure}[htb]
\centering 
\scalebox{0.4}{\includegraphics{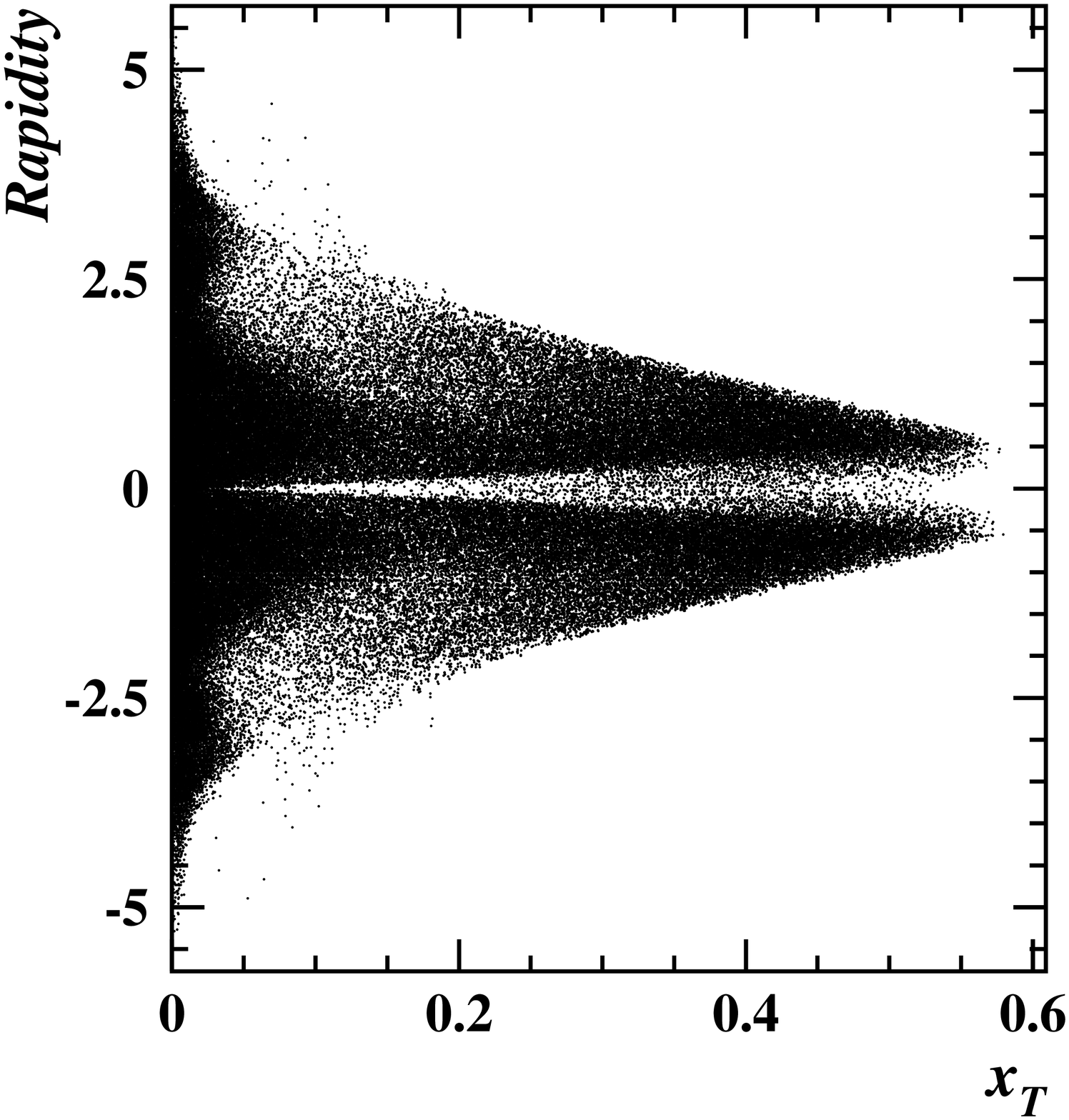}} 
\caption{The particle-number distribution versus both the rapidity and the transverse momentum defined with respect to the thrust axis. The two dark wings of the swallowtail correspond to two $\bar {c}$ jets. The dark area around $x_T=0$ corresponds to the third jet, including $\Xi_{cc}/T_{cc}$.}
\label{pty} 
\end{figure}

As a more complete demonstration of the properties of symmetric three-jet events, we present the particle-number distribution versus both the rapidity and the transverse momentum.
The swallowtail structure in Fig. \ref{pty} clearly demonstrates the string effect.
The three jets are well separated on the contour of the particle rapidity and the transverse momentum.
Once the $\Xi_{cc}$ or $T_{cc}$ is observed and a 
large number of events is accumulated, this kind of effect can be convincingly identified at $e^+e^-$ colliders. The color-flow method can also be extended to study resonant heavy-colored-particle production at the LHC and the ILC.

\section{Summary}

In this paper, we investigate a special color connection of the four-heavy-quark system and the corresponding hadronization effects in
$e^{+}e^{-}$ annihilation. We point out that in addition to the normal color
structure with two color singlets, which independently fragment into hadrons
in the popular models, there exists another kind of color structure,
as shown in Eq. (\ref{q2qq}), which must be taken into account
in some cases, \textit{e.g.}, $\Xi_{cc}$ and $T_{cc}$ production. For this kind
of color connection, the hadronization of the four quarks should
be treated as an entire system, \textit{i.e.}, they interact with one another during the hadronization process. We discuss the hadronization
for this kind of color connection and investigate the triggers for
the corresponding effects. The most significant one is the doubly heavy
baryon/tetraquark production in three-jet-like events. Furthermore,
for symmetric three-jet events, the string effects of
the special particle flow can be used to track the color connection more
precisely. $e^{+}e^{-}$ colliders are good arena to study various
color connections of final multiparton systems. There exist several $e^{+}e^{-}$ collider projects, \textit{e.g.}, the ILC, the CLIC, Higgs
factories and $Z^{0}$ factories. All these projects can incorporate the
Giga-Z project and are excellent opportunities for experimentally verifying our suggestions.

\section*{Acknowledgments}

This work is supported in part by NSFC (11275114, 11047029, 10935012) and 
the Natural Science Foundation of Shandong Province. The authors thank
all members of the Theoretical Particle Physics Group of Shandong
University for their helpful discussions.


\end{document}